\documentclass[aip,pop,graphicx,amsmath,amssymb,reprint]{revtex4-1}
\usepackage{dcolumn}
\usepackage{bm}
\usepackage{units}
\usepackage{epsfig}
\usepackage{color}
\usepackage{todonotes}
\draft 

\begin{document}


\title[]{Nonthermal ion acceleration by the kink instability in nonrelativistic jets}




\author{E. P. Alves}
\email{epalves@slac.stanford.edu}
\affiliation{
High Energy Density Science Division, SLAC National Accelerator Laboratory, Menlo Park, CA 94025, USA
}

\author{J. Zrake}
\affiliation{
Physics Department and Columbia Astrophysics Laboratory, Columbia University, 538 West 120th Street, New York, NY 10027
}

\author{F. Fiuza}
\email{fiuza@slac.stanford.edu}
\affiliation{
High Energy Density Science Division, SLAC National Accelerator Laboratory, Menlo Park, CA 94025, USA
}

\date{\today}

\begin{abstract}
We investigate the self-consistent particle acceleration physics associated with the development of the kink instability (KI) in nonrelativistic, electron-ion plasma jets. Using 3D fully kinetic particle-in-cell (PIC) simulations, we show that the KI efficiently converts the initial toroidal magnetic field energy into energetic ions. The accelerated ions form a nonthermal power-law tail in the energy spectrum, containing $\simeq10\%$ of the initial magnetic field energy, and with the maximum ion energy extending to the confinement energy of the jet. We find that the ions are efficiently accelerated by the concerted action of the motional electric field and highly tangled magnetic field that develop in the nonlinear phase of the KI: fast curvature drift motions of ions across magnetic field lines enable their acceleration along the electric field. We further investigate the role of Coulomb collisions on the ion acceleration efficiency, and identify the collisional threshold above which nonthermal ion acceleration is suppressed. Our results reveal how energetic ions may result from unstable nonrelativistic plasma jets in space and astrophysics, and provide constraints on the plasma conditions required to reproduce this acceleration mechanism in laboratory experiments.
\end{abstract}


\maketitle 

\section{Introduction}

Jets of plasma threaded by magnetic fields are ubiquitous in space and astrophysical environments. Throughout their evolution, these jets can be subject to current-driven magnetohydrodynamic (MHD) instabilities, the most common being the sausage ($m=0$) and kink ($m=1$) modes \cite{Bateman1978}. The development of such instabilities is believed to be a candidate mechanism to explain the conversion of the jet's magnetic energy into energetic particles and radiation in a wide range of space physics and astrophysics scenarios. These include the flaring activity and energetic particle generation in jets emanating from the solar atmosphere \cite{Pariat2015,Bucik2018a,Bucik2018b}, the synchrotron emission from bright knots and cosmic ray acceleration in jets from active galaxies \cite{Giannios2006,Bromberg2016,Alves2018}, among others. These instabilities are also associated with violent disruptions and particle acceleration in laboratory fusion devices, such as Tokamaks \cite{Zakharov2012} and Z-pinches \cite{Ryutov2000,Haines2011,SuzukiVidal2013}. Yet, the detailed physics underlying how the development of such instabilities in jets accelerates particles remains poorly understood.

MHD simulations play an important role in understanding where and when sausage and kink modes develop over the course of the formation and propagation of astrophysical \cite{Bromberg2016,Tchekhovskoy2016} and laboratory \cite{Ciardi2007} plasma jets, and how their nonlinear development impacts the structural integrity of jets \cite{Mizuno2011a}. However, the kinetic physics that underpins the acceleration of particles is not captured by MHD models. Simulations of test particle dynamics in the fields computed by MHD simulations have shown nonthermal acceleration \cite{Ripperda2017} but do not provide a self-consistent picture of the injection mechanisms or feedback of energetic particles on the plasma fields. Fully kinetic simulations are necessary to address this central problem in a self-consistent manner. Using 3D PIC simulations, we have recently demonstrated the self-consistent acceleration of nonthermal particles resulting from the development of the KI in relativistic magnetized pair plasma and pair-proton jets\cite{Alves2018}. These simulations revealed that the KI provides a viable means of producing radiating nonthermal particle distributions and of accelerating ultra-high-energy cosmic rays in the jets of active galactic nuclei. These findings encourage further exploration of the KI's ability to efficiently accelerate particles in nonrelativistic electron-ion plasma jets of relevance to a range of space and laboratory conditions.

In this paper, we report on 3D fully kinetic simulations of the KI in nonrelativistic, electron-ion plasma jets and discuss the associated particle acceleration physics. We demonstrate that the nonlinear development of the KI in collisionless plasma results in the efficient dissipation of the jet's magnetic field energy, which is preferentially transferred to high-energy ions. Approximately  $10\%$ of the initial magnetic field energy is converted into nonthermal ions. The ion energy spectrum develops a power-law tail $\propto \varepsilon^{-p}$, with $p$ approximately fixed at $\simeq2.6$ over the parameter range explored in this work. We discuss the ion acceleration mechanism and how it depends on the size and magnetization of the jet, and how the mass ratio between electrons and ions dictates the energy partition between species. In addition, we explore how the plasma collisionality impacts particle acceleration, which is relevant for a variety of laboratory experiments. Previous experiments using plasma guns \cite{HsuBellan2002, HsuBellan2003}, modified plasma Z-pinch platforms \cite{Lebedev2002,Lebedev2005,SuzukiVidal2013,Lebedev2019} and high-power lasers \cite{CKLi2016} have successfully studied MHD instabilities in jets. In some cases, the observation of nonthermal ions has been reported \cite{SuzukiVidal2013}, yet without a clear understanding of the underlying acceleration mechanism. Our work helps establish the conditions for which the present acceleration mechanism can be studied in laboratory experiments.

This paper is organized as follows. In Section \ref{sec:sim_setup}, we describe the initial physical configuration of the plasma jet and our simulation setup. We then discuss the evolution of the electric and magnetic fields produced by the KI and the associated particle acceleration dynamics in the collisionless regime in Section \ref{sec:collisionless_regime}. The effects of Coulomb collisions on the particle acceleration dynamics are discussed in Section \ref{sec:collisional_regime}, and conclusions are drawn in Section \ref{sec:conclusions}.

\section{Simulation setup}
\label{sec:sim_setup}
 We simulate the particle acceleration dynamics associated with the development of the KI in a nonrelativistic electron-ion plasma jet using the 3D PIC code OSIRIS 4.0 \cite{Fonseca2002,Fonseca2008}. We consider a purely toroidal magnetic field profile of the form $B_\phi(r)=B_0\frac{r}{R}e^{1-r/R}$ where $B_0$ is the peak amplitude of the magnetic field and $R$ is the characteristic radius of the jet. The current density $\mathbf{J} = c/4\pi \nabla\times\mathbf{B}$ is supported by symmetrically streaming electrons and ions along the jet axis. The thermal pressure profile of the plasma ($P$) is chosen to achieve hydromagnetic equilibrium, $\nabla P = \mathbf{J}\times\mathbf{B}$; we consider plasma with uniform number density $n_e = n_i = n_0$ and nonuniform temperature profile with $T_e = T_i= T(r)$, such that $P(r)=2n_0k_BT(r)$. Note that we have tested different initial conditions with jet current being carried solely by the electrons and no net flow of ions, and have found that the particle acceleration dynamics remains unaffected. This is to be expected since the particle thermal velocities are significantly higher than the flow velocities required to support the electric current.

The jet's initial conditions can thus be characterized by the following parameters. A dimensionless measure of the jet's magnetization $\omega_{ci}/\omega_{pi}$, where $\omega_{ci}= eB_0/m_ic$ is the ion cyclotron frequency and $\omega_{pi}= \sqrt{4\pi n_0e^2/m_i}$ is the ion plasma frequency; the dimensionless jet radius $R/d_i$, with $d_i = c/\omega_{pi}$ being the ion inertial length; and the ion to electron mass ratio $m_i/m_e$. Note that due to pressure balance $n_0k_BT_0 = B_0^2/4\pi$, we have that the typical Larmor radius of thermal ions at the core of the jet is given by $\bar{\rho_i}=\sqrt{m_i k_BT_0} c/eB_0 = d_i$. 

We explore jet magnetizations in the range $0.08\leq \omega_{ci}/\omega_{pi} \leq 0.25$ (corresponding to nonrelativistic magnetic energy densities $\sigma_i \equiv (\omega_{ci}/\omega_{pi})^2$ in the range $0.006-0.06$), and we simulate jet radii in the range $R/d_i = 5 - 10$. Given the high computational cost of simulating realistic ion to electron mass ratios, we perform simulations with reduced mass ratios $m_i/m_e$ between $4-36$. By progressively increasing mass ratio we uncover how the particle acceleration physics scales with $m_i/m_e$, allowing us to infer the behavior for realistic mass-ratio conditions.

 The dimensions of our simulation domain are $20R\times20R\times10R$, with the jet placed at the center of the domain and oriented along $\mathbf{\hat{z}}$. The grid resolution is chosen to resolve the the gyroradius of thermal electrons at the core of the jet $\bar{\rho_e} = \bar{\rho_i} \sqrt{m_e/m_i}$ with $> 2$ points, and we use $12$ particles per cell per species with quadratic splines for the particle shapes. Finer spatial resolutions and more particles per cell were tested to ensure numerical convergence. Periodic boundary conditions are imposed in all directions; the transverse ($\mathbf{\hat{x}}$ and $\mathbf{\hat{y}}$) dimensions of the domain are sufficiently large to avoid artificial recirculation of particles during the time-scale of interest. We follow the evolution of the system until the particle acceleration dynamics terminates, which occurs at $\simeq 25-30~R/v_A$ ($v_A = B_0/\sqrt{4\pi n_0 m_i}$ is the characteristic Alfv\'en speed).

\begin{figure*}[t!]
\begin{center}
\includegraphics[width=\textwidth]{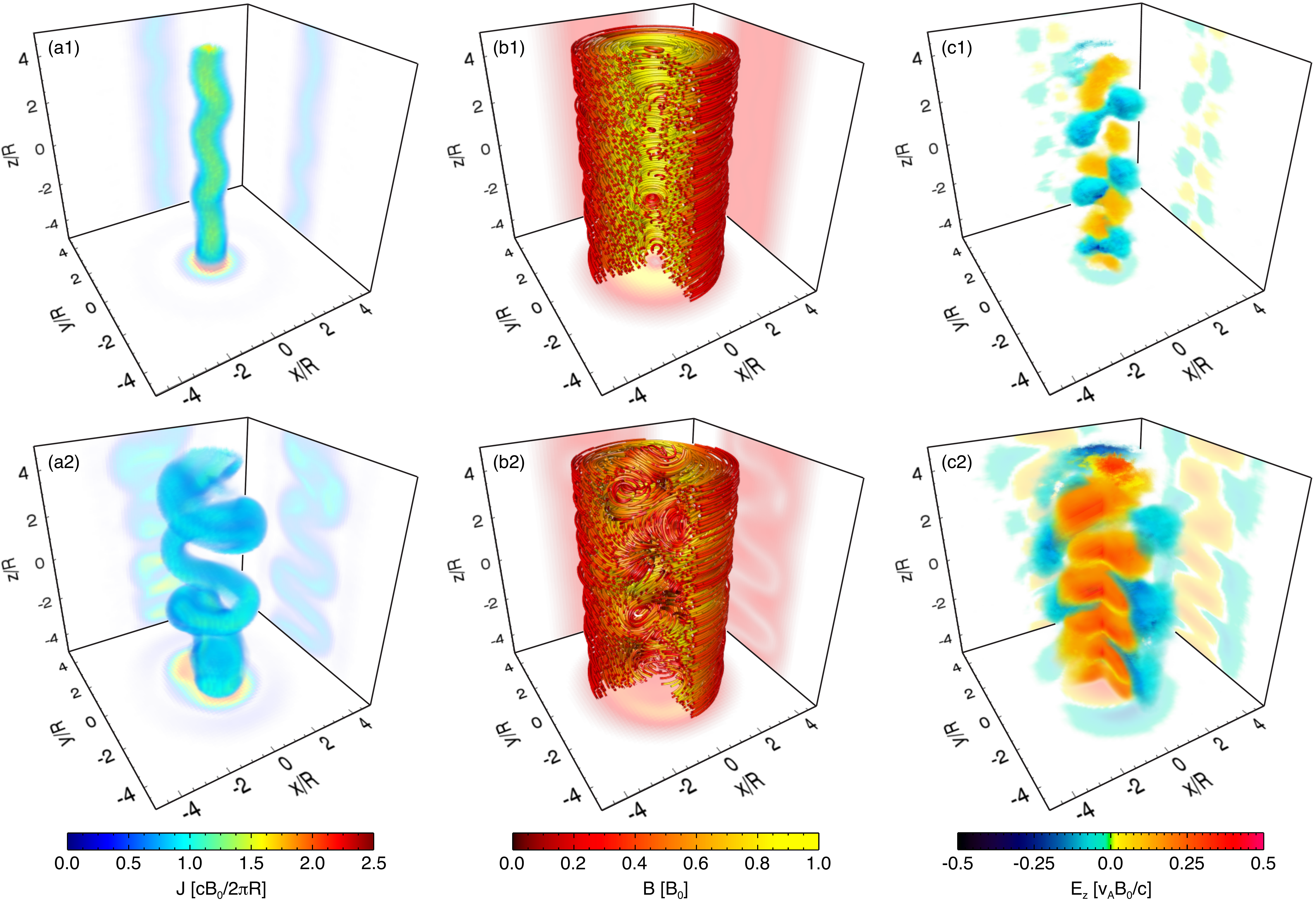}
\caption{Development of the kink instability in a collisionless, nonrelativistic electron-ion jet (with $\omega_{ci}/\omega_{pi}=0.125$, $R/d_i=5$ and $m_i/m_e=16$). (a) Current density, (b) magnetic field lines, and (c) axial electric field, taken at times (1) $v_A t/R = 8$ and (2) $v_A t/R = 12.5$. These times correspond to the linear and nonlinear stages of the kink instability. Note that a quarter of the simulation box has been removed in (b1), (b2), and (c2) to reveal the inner field structure of the jet.
}
\end{center}
\end{figure*}

\section{Particle acceleration in collisionless electron-ion jets}
\label{sec:collisionless_regime}

\subsubsection{Structure and evolution of the KI-induced E and B fields}
The general dynamics of the KI in collisionless electron-ion plasma is illustrated in Figure 1 with the case of $\omega_{ci}/\omega_{pi} = 0.125$ ($\sigma_i = 0.016$), $m_i/m_e = 16$ and $R/d_i = 5~(=R/\bar{\rho_i})$; all other simulations explored in this work exhibited similar behavior. The onset of the KI is triggered by small transverse displacements of the jet about its equilibrium. Those displacements produce imbalances in the magnetic pressure across the jet, inducing motions that reinforce the initial distortion. This ultimately manifests as a growing helical modulation of the jet structure (left column of Figure 1), with a wavelength $\simeq 2.5 R$ and developing at a rate $\sim v_A/R$, in agreement with linear MHD theory \cite{Bateman1978}.

The radial motions of the magnetized plasma jet give rise to an inductive electric field $\mathbf{E} = -(\mathbf{v}/c)\times\mathbf{B}$, which at early times is also shown to have a helical and harmonic structure in Figure 1 (c1). Similar to the behavior found in the relativistic pair plasma regime \cite{Zenitani2005, Alves2018}, the nonlinear distortions of the jet (distortions comparable to the jet radius $R$) result in the formation a coherent structure in the axial component of the electric field $E_z$ [Figure 1 (c2)]; the mean amplitude of the electric field along the jet axis is found to be $\langle E_z\rangle_z \simeq 0.3 v_AB_0/c$. The nonlinear distortions of the jet current density also result in a highly tangled magnetic field structure [Figure 1 (b2)]. It is this configuration of electric and magnetic fields that mediates the efficient conversion of magnetic energy ($\simeq 65\%$ of initial magnetic energy $\varepsilon_B(0)$) into plasma kinetic energy [Figure 2 (a)]. This efficient accelerating field structure persists during the transit time of the kink perturbations across the diameter of the jet, $\tau_{KI} \sim 2R/v_{KI} = 6R/v_A$. After this period, the fields decay and particle acceleration ceases.

\begin{figure*}[t!]
\begin{center}
\includegraphics[width=\textwidth]{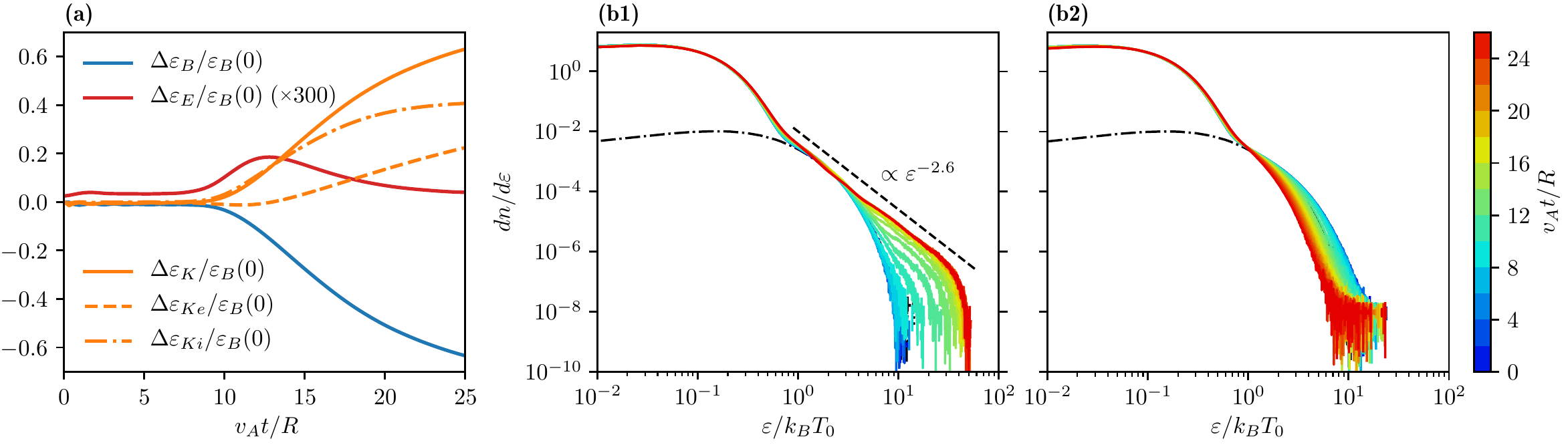}
\caption{(a) Temporal evolution of the change in magnetic ($\Delta\varepsilon_B$), electric ($\Delta\varepsilon_E$) and particle kinetic energies ($\Delta\varepsilon_K$) integrated over the simulation domain. The separate evolution of the change in electron ($\Delta\varepsilon_{Ke}$) and ion kinetic energies ($\Delta\varepsilon_{Ki}$) are also shown. Temporal evolution of the ion (b1) and electron (b2) energy spectra, integrated over the simulation domain. Note that the double hump structure of both ion and electron spectra at $t=0$ corresponds to the contributions of the warm jet plasma (that balances the hoop stress of the magnetic field), and the cold ambient plasma outside the jet (also providing confinement of the jet). The initial energy distribution of the warm jet plasma is represented by the black dash-dotted curves in (b1) and (b2).
}
\end{center}
\end{figure*}

\subsubsection{Efficient acceleration of nonthermal ions}
As shown in Figure 2 (a), the dissipated magnetic energy is preferentially transferred to the jet ions, gaining $\simeq2\times$ as much energy as the electron population. More interestingly, this process results in the acceleration of non-thermal ions, forming a high-energy power-law tail in the ion energy spectrum $\propto\varepsilon^{-p}$ with $p=2.6$ [Figure 2 (b1)]. We find that $\simeq 10\%$ of the total initial magnetic field energy is transferred to nonthermal ions with $\varepsilon>5K_BT_0$. We further verify that ions are accelerated up to the confinement energy of the jet $\varepsilon_\mathrm{conf} \simeq (eB_0R)^2/2m_ic^2 = k_BT_0(R/d_i)^2$ (valid for subrelativistic ion energies), i.e. the energy beyond which the ion Larmor radius $\rho_i$ exceeds the system size $R$. In the astrophysical context, this limiting energy is also known as the Hillas energy or the Hillas constraint \cite{Hillas1984}. For the simulated parameters, we find that the cutoff energy of the ion spectrum occurs at $\varepsilon \simeq 40 k_BT_0 = 1.6 \varepsilon_\mathrm{conf}$. This is consistent with the maximum energy gain of an ion accelerating freely in the coherent axial electric field structure $\langle E_z \rangle$ during the time-scale $\tau_{KI}$, $\varepsilon_\mathrm{max} \simeq (e \langle E_z\rangle \tau_{KI})^2/2m_i \sim \varepsilon_\mathrm{conf}$.

The jet electrons, however, are not accelerated to nonthermal energies by the KI [Figure 2 (b2)]. In fact, the electrons within the jet (within $r<R$) cool down as shown by the receding spectral tail in Figure 2 (b2). It is the electrons at the periphery of the jet ($r\gtrsim R$) that absorb a fraction of the dissipated magnetic energy, but are not accelerated efficiently to many times their initial thermal energy. The difference between the acceleration dynamics of ions and electrons will be further discussed in the next Section.

\begin{figure*}[t]
\begin{center}
\includegraphics[width=\textwidth]{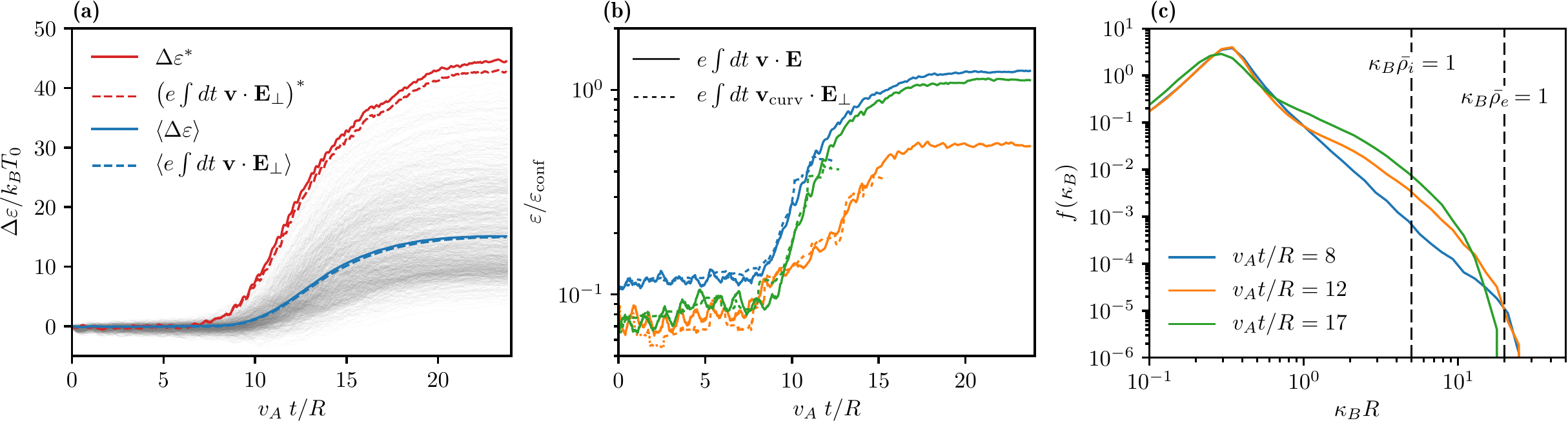}
\caption{ (a) Evolution of the energy gain $\Delta\varepsilon$ of a representative sample of $2000$ ions that reach the nonthermal power-law tail (grey curves), self-consistently accelerated via the development of the KI. The ion with the highest energy gain of the sample and the mean energy gain of the sample are represented by the solid red and blue curves, respectively. The evolution of the work performed by the motional electric field $\mathbf{E_\perp}$ is represented by the dashed red and blue curves for comparison. (b) The energy gain of a random subset of $3$ ion trajectories (from the original sample of $2000$ trajectories), represented by the solid curves, is compared to the work performed by $\mathbf{E_\perp}$ along the curvature drift trajectory of their guiding centers (dashed curves); the dashed curves are interrupted when the ion Larmor radii become a significant fraction of the jet radius, resulting in the break down of the guiding center drift description of the ion trajectories. (c) Evolution of the distribution of magnetic field curvature $\kappa_B$ in the simulated domain at $v_At/R = 8$, $12$ and $17$. The vertical dashed lines mark the points where the magnetic curvature equals the Larmor scale of thermal ions ($\kappa_B\bar{\rho_i} =1$) and electrons ($\kappa_B\bar{\rho_e} =1$).
}
\end{center}
\end{figure*}

\subsubsection{Ion acceleration mechanism}

We find that the mechanism by which nonthermal ions are accelerated is similar to that identified in the regime of relativistic pair plasma \cite{Alves2018}. By following the trajectories of a representative sample of nonthermal ions [grey curves in Figure 3(a)], we have verified that their acceleration is due to the work of the motional electric field $\mathbf{E} = -\mathbf{v}\times\mathbf{B} \equiv \mathbf{E_\perp}$. This is shown for both the ion that attained the highest energy in the sample (red curve) and the mean energy gain of the sample (blue curve) in Figure 3 (a). This indicates that non-ideal electric fields ($\mathbf{E_\parallel}$), commonly associated with reconnecting current layers, have a negligible role in the acceleration dynamics of ions.

The efficient acceleration of ions by the motional electric field $\mathbf{E_\perp}$ implies that ions must move efficiently transverse to the local magnetic field. Indeed, we find that the highly inhomogeneous and highly tangled structure of the magnetic field that develops in the nonlinear phase of the KI [Figure 1 (b2)] facilitates the displacement of ions across magnetic field lines via guiding center drift motions. In particular, we find that curvature drift motions of the ions play a dominant role. This is verified in Figure 3 (b), which shows that the energy gain of a random subset of $3$ particle tracks (from the same sample) is well described by the work performed by the motional electric field along their guiding center curvature drift trajectories, i.e. $\Delta \varepsilon \simeq e\int \mathbf{v}_\mathrm{curv} \cdot \mathbf{E_\perp}$, where $\mathbf{v}_\mathrm{curv}=v_\parallel^2\mathbf{B}\times\kappa_B/\omega_{ci}|\mathbf{B}|$ is the curvature drift velocity and $\mathbf{\kappa_B} = \mathbf{B}\cdot\nabla\mathbf{B}/|B|^2$ is the magnetic curvature vector field. The guiding center description of the ion trajectories breaks down when they achieve a large fraction of the confinement energy of the jet, $\simeq 0.3\varepsilon_\mathrm{conf}$. Beyond these energies, ions become effectively unmagnetized, moving nearly freely along the motion electric field in meandering type orbits until they escape the acceleration region, or until the accelerating fields decay.

Figure 3 (c) show the evolution of the distribution of magnetic curvature $\kappa_B$ during the nonlinear development of the KI. Enhancement in the distribution of magnetic curvature is observed in between the scales of the jet radius $\kappa_B R\simeq 1$ and Larmor radius of thermal electrons $\kappa_B \bar{\rho_e}\simeq 1$. However, this enhancement is significantly larger at the Larmor scale of thermal ions when compared to that of thermal electrons. This result indicates that thermal ions are more likely to encounter magnetic field curvature that approaches the scale of their Larmor radii than electrons, allowing them to be more easily injected into a rapid accelerating phase via curvature drifts. Moreover, ions can become locally unmagnetized (when they experience $\kappa_B \rho_i \gtrsim 1$) with higher probability than electrons. It is likely that this distribution of magnetic curvature is key to explaining the difference in acceleration efficiency between the two species. A more in detailed analysis of the electron dynamics and acceleration efficiency for different initial conditions will be the subject of future work.

\begin{figure}[h]
\begin{center}
\includegraphics[width=0.75\columnwidth]{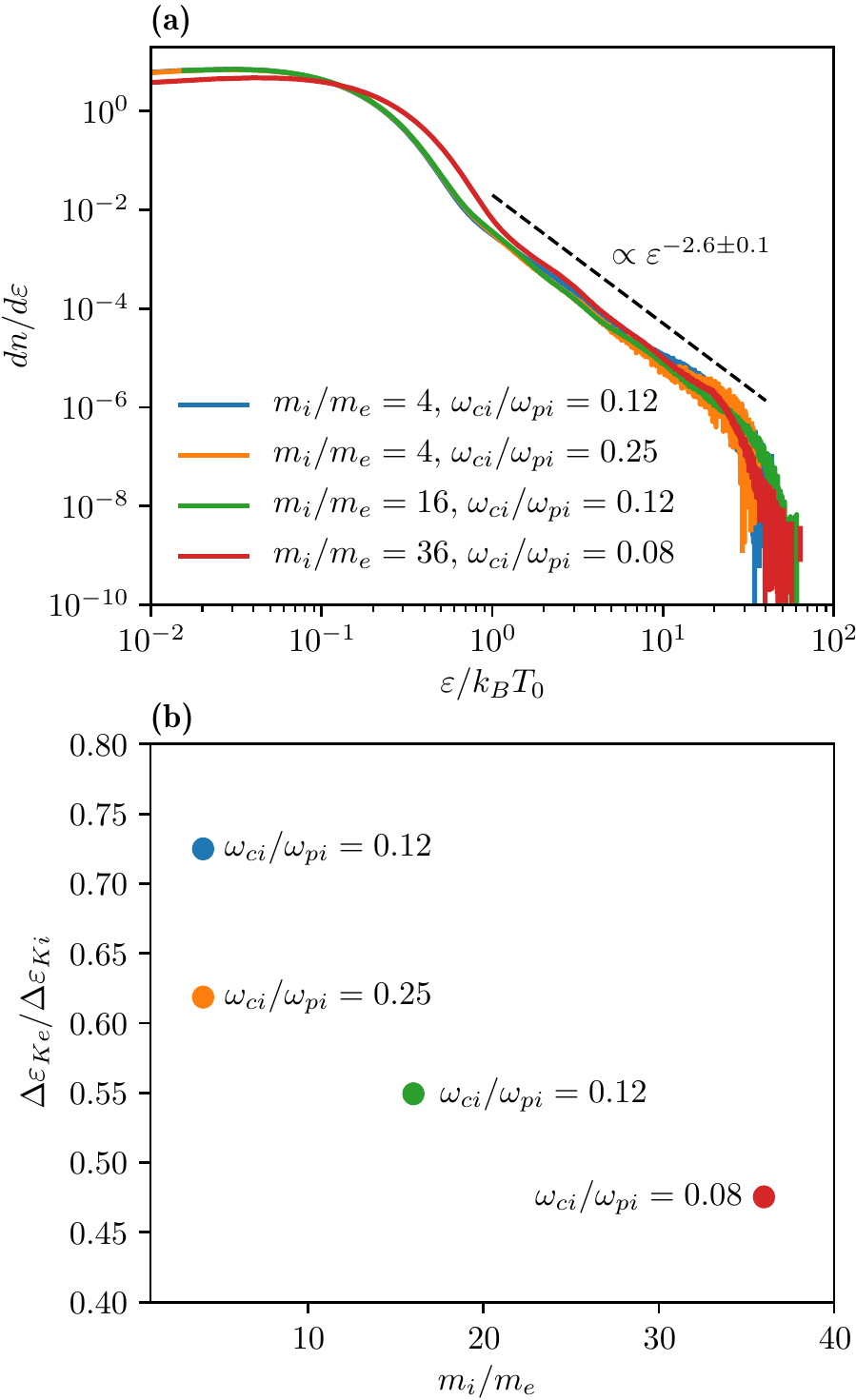}
\caption{(a) Energy spectrum of accelerated ions for varying ion to electron mass ratio ($m_i/m_e$) and jet magnetization ($\omega_{ci}/\omega_{pi}$) at the fixed dimensionless jet radius $R/d_i=5$. Note that the energy scale is normalized to the thermal energy $k_BT_0$, which varies with jet magnetization. Further note that the slight energy-shift observed in the spectrum of the $m_i/m_e=36$ case was due to the use of a slightly higher initial plasma temperature that raised the plasma thermal pressure uniformly throughout the simulation domain (without affecting the initial hydromagnetic equilibrium). This slight change in the initial conditions did not influence the development of the high-energy nonthermal component of the spectrum. (b) Dependence of the electron to ion energy gain ($\Delta\varepsilon_{Ke}/\Delta\varepsilon_{Ki}$) on the mass ratio. Matching colors between the spectra in (a) and the points in (b) correspond to the same simulated parameters.}
\end{center}
\end{figure}

\subsubsection{Particle acceleration dependence on magnetization ($\omega_{ci}/\omega_{pi}$), system size $R/\bar{\rho_i}$ and mass ratio $m_i/m_e$}

We have performed a set of simulations of varying magnetization ($0.08\leq \omega_{ci}/\omega_{pi}\leq0.25$), system size ($5\leq R/d_i\leq10$) and ion to electron mass ratio ($4\leq m_i/m_e \leq 36$) to probe how these parameters influence the particle acceleration dynamics of the KI. For the range of parameters explored in this work, we have verified that all simulations exhibit identical growth and saturation of the KI upon normalizing space and time to $R$ and $v_A/R$, respectively. We systematically observe the conversion of $\simeq 65\%$ of the initial magnetic energy in the system into plasma kinetic energy after $25~R/v_A$ in all our simulations. The magnitude of the accelerating electric field structure that develops in the nonlinear phase of the KI is also similar across our simulations when normalized to the characteristic value of $v_A/c B_0$ ($\langle E_z\rangle_z \simeq 0.3 v_AB_0/c$).

We observe the persistent acceleration of nonthermal ions in all our simulations, while electrons are always found to remain thermal. The spectra of accelerated ions for varying $m_i/m_e$ and $\omega_{ci}/\omega_{pi}$ at fixed system size $R/d_i = 5$ are overlaid in Figure 4 (a). Interestingly, we find that the power-law index of the nonthermal tail remains approximately constant at $-2.6\pm0.1$ for varying $m_i/m_e$ and $\omega_{ci}/\omega_{pi}$ (within the explored parameter range), and thus these results may be extrapolated to realistic mass ratios. We further verify in all simulations that the nonthermal ion tail extends up to the confinement energy of the jet $\varepsilon_\mathrm{conf} = k_BT_0(R/d_i)^2$ [Figure 4 (a)]. For a fixed normalized jet radius $R/d_i =5$, we find that $\varepsilon_\mathrm{conf}/k_BT_0 = (R/d_i)^2 = 25$, which is independent of the ion to electron mass ratio and jet magnetization, as seen in Figure 4 (a). In all simulated cases, we find that the cutoff energy of the spectrum is $\varepsilon_\mathrm{max}/k_BT_0 \simeq 40 = 1.6 \varepsilon_\mathrm{conf}/k_BT_0$.

While the fraction of initial magnetic energy that is dissipated is approximately constant in all our simulations, the ratio of the electron to ion energy gain ($\Delta\varepsilon_{Ke}/\Delta\varepsilon_{Ki}$) is shown to depend on both the mass ratio and the magnetization, as illustrated in Figure 4 (b). We observe that an increasing fraction of the dissipated magnetic energy is transferred to the ions both with increasing mass ratio and increasing magnetization. We note that a similar dependence of $\Delta\varepsilon_{Ke}/\Delta\varepsilon_{Ki}$ on $m_i/m_e$ has been reported in 3D PIC simulations of relativistic magnetized plasma turbulence\cite{Zhdankin2019}.

\begin{figure*}[t!]
\begin{center}
\includegraphics[width=\textwidth]{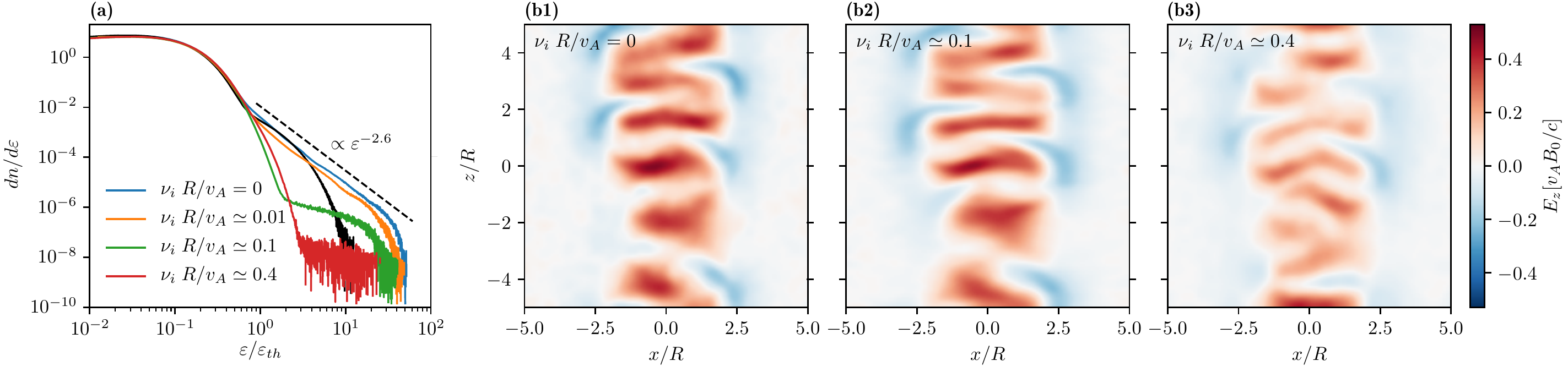}
\caption{Impact of Coulomb collisions on ions acceleration efficiency. (a) Energy spectrum of accelerated ions for varying ion-collision frequency ($\nu_i$); the jet parameters are $R/d_i=5$, $\omega_{ci}/\omega_{pi} = 0.125$ and $m_i/m_e = 16$. The ion collision frequency is varied by artificially varying the Coulomb logarithm. The solid black curve represents the initial ion spectrum, before the development of the KI. Panels (b), (c) and (d) are cross-sections of the of the $\mathbf{\hat{z}}$-component of the electric field at $v_At/R\simeq12.5$ for the different collisionalities $\nu_iR/v_A=0$, $0.1$ and $0.4$ respectively.}
\end{center}
\end{figure*}

\section{Role of Coulomb collisions on ion acceleration efficiency}
\label{sec:collisional_regime}

There is a significant experimental effort to study the MHD stability of plasma jets in the laboratory in conditions relevant for astrophysical environments \cite{HsuBellan2002,HsuBellan2003, Lebedev2002,Lebedev2005,SuzukiVidal2013,CKLi2016,Lebedev2019}. However, laboratory-produced plasma jets are generally far more collisional than those naturally occurring in space or astrophysical settings. Thus, in order to assess if the ion acceleration mechanism discussed in the previous Section can be studied in laboratory plasma experiments, it is important to consider the effects of Coulomb collisions. One naturally anticipates that if the jet plasma is too collisional, nonthermal particle distributions will not be allowed to develop as they will rapidly relax to thermal equilibrium.

We have investigated the impact of a finite Coulomb collision frequency on the particle acceleration dynamics of the KI, with the aim of determining the collisional threshold above which nonthermal particle acceleration is suppressed. We include the physics of binary Coulomb collisions (the Spitzer-Harm model for weakly collisional plasmas) in our PIC simulations using a Monte-Carlo approach \cite{Takizuka1977,Nanbu1998,Sentoku2008,Perez2012}. This method randomly pairs particles locally within a cell, and scatters their momenta such that energy and momentum are conserved on each collision (equal particle weights are used). We use this method to model electron-ion and ion-ion collisions in our simulations of the KI, and we vary the collision frequency by artificially varying the Coulomb logarithm. This way, all other physical parameters of the system can be held fixed (plasma density, temperature, magnetic field strength) and the effect of Coulomb collisions can be varied independently.

We have examined the effect of Coulomb collisions on the illustrative case explored in the previous Section, i.e. a jet with $\omega_{ci}/\omega_{pi}=0.125$, $m_i/m_e=16$ and $R/d_i=5$ (note that $R/d_i =5$ is close to the jet radii produced in recent experiments \cite{Lebedev2005, SuzukiVidal2013,CKLi2016}). We progressively increased the ion collision frequency $\nu_i$ relative to the dynamical time the KI (which also corresponds to the acceleration time-scale of the KI) from $\nu_i R/v_A =0$ to $0.4$; note that $\nu_i R/v_A$ can also be conveniently interpreted as the ratio of the jet radius to the ion mean-free-path, $\nu_i R/v_A\simeq R/\lambda_i$, since $v_A\simeq\sqrt{k_BT_0/m_i}$ (as imposed by the initial equilibrium configuration). The resulting spectra of accelerated ions are shown in Figure 5 (a) for varying $\nu_i$. We find that the spectrum of accelerated ions remains unaffected for $\nu_iR/v_A \lesssim 0.01$, relative to the collisionless case: the slope of the nonthermal tail, the number of nonthermal particles and the maximum energy reached remain unchanged by Coulomb collisions below $\nu_iR/v_A \simeq 0.01$. At $\nu_iR/v_A \simeq 0.1$, the ion collision frequency is no longer negligible compared to the acceleration time-scale of the KI. As ions gain energy through the KI-induced electric field, they also lose energy through collisions and heat up the background particles. This effect leads to a strong reduction in the acceleration efficiency of nonthermal ions as shown in Figure 5 (a): the number of nonthermal particles is strongly reduced and the slope of the tail of the distribution is hardened. At $\nu_iR/v_A \simeq 0.4$, nonthermal ion acceleration is nearly fully suppressed.

Note that as the collision frequency was varied, no significant changes were observed in the morphology or temporal evolution of the KI-induced electric and magnetic fields relative to the collisionless regime. This is illustrated Figures 5 (b1-3) by the snapshots of the cross-section of $E_z$ for varying collisionality. These snapshots were all taken at $v_At/R\simeq12.5$, revealing that the temporal development and spatial structure of the KI-induced electric field is nearly unaffected by collisions; only in the highly collisional case of $\nu_iR/v_A\simeq0.4$ does the amplitude of $E_z$ become noticebly lower, mainly as a consequence of the increased magnetic diffusivity that lowers the peak amplitude of the magnetic field, and hence of the motional electric field.

The laboratory plasma jets produced by the radial wire array Z-pinch \cite{Lebedev2005}, are characterized by low-temperature and high-Z plasmas, achieving $R/\lambda_i \sim 10^6$. The acceleration of nonthermal ions by the mechanism reported here is thus expected to be significantly suppressed within the dense body of the jet. However, high-energy ions have indeed been reported resulting from the development of MHD instabilities in the jets produced by this platform\cite{SuzukiVidal2013}. It is likely these high-energy ions are accelerated outside of the dense jet, in the low-density ambient plasma where the collisionality is reduced. Our simulations have not considered the role of a low-density plasma background surrounding the jet, which could allow ambient particles to interact with the fields produced by the KI of the jet while remaining in a weakly collisional environment. Indeed we do see in our simulations that particles outside of the jet, within $R<r<2R$, do interact with the KI-induced fields and absorb a significant fraction of the dissipated magnetic energy. It is therefore possible that the mechanism reported here can participate in the acceleration of nonthermal particles in the low-density plasma surrounding the dense jet. A more detailed exploration of our work in the conditions of these experiments will be pursued in the near future.

Laser-driven high-energy-density plasmas are more likely to produce weakly collisional jets in the laboratory. Recent laser-driven plasma experiments produced near keV-temperature plasma jets, with $R/\lambda_i \sim 10^2$, aimed at investigating the development of the KI in astrophysically relevant conditions \cite{CKLi2016}. Future experiments using a more energetic laser drive may be able to produce jets with $R/\lambda_i \lesssim 0.1$, making it possible to directly probe the efficient particle acceleration physics of the KI within the body of the jet, as described in this work.

\section{Conclusions}
\label{sec:conclusions}

We have shown via 3D PIC simulations that the development of the KI in nonrelativistic, electron-ion plasma results in the efficient acceleration of nonthermal ions. Approximately $10\%$ of initial magnetic energy is transferred into nonthermal ions over the course of a few $10$'s of dynamical times of the KI ($R/v_A$), forming a power-law tail in the energy spectrum with index $\simeq 2.6$. We showed that the power-law index of the nonthermal tail remains nearly constant over the range of jet magnetizations explored in our work, and that the maximum ion energy systematically reaches the confinement energy of the jet $\varepsilon_\mathrm{conf}/k_BT_0 \simeq (R/d_i)^2$.

The ion acceleration mechanism is similar to the mechanism found to operate in relativistic pair plasma jets \cite{Alves2018}. The nonlinear development of the KI produces a coherent motional electric field along the axis of the jet that is embedded in a highly tangled magnetic field. Ions experience fast curvature drift motions in the highly tangled magnetic field that permit their efficient displacement along the motional electric field, and hence their efficient acceleration. Our results indicate that this is a viable mechanism to explain efficient acceleration of nonthermal protons in nonrelativistic jets in space and astrophysical environments. 

The similarity between the particle acceleration physics in the relativistic and nonrelativistic regimes of the KI should also motivate the development of laboratory experimental platforms capable of investigating the particle acceleration mechanisms relevant to astrophysical jets. Our work indicates that in order to study this acceleration mechanism in the laboratory, the plasma collisionality needs to be significantly reduced, such that the ion mean free path becomes greater than $10\times$ the jet radius. These conditions can likely be produced in laser-driven high-energy-density plasma experiments \cite{Huntington2015}.

\begin{acknowledgments}
This work was supported by the U.S. Department of Energy SLAC Contract No. DE-AC02-76SF00515, by the U.S. DOE Office of Science, Fusion Energy Sciences under FWP 100237, and by the U.S. DOE Early Career Research Program under FWP 100331. The authors acknowledge the OSIRIS Consortium, consisting of UCLA and IST (Portugal) for the use of the OSIRIS 4.0 framework and the visXD framework. Simulations were run on Mira (ALCF) through an ALCC award.
\end{acknowledgments}

\bibliography{scibib}

\end{document}